\documentclass{ws-mpla}
\usepackage[super]{cite}
\usepackage{graphicx}

\renewcommand{\v}[1]{\boldsymbol{#1}}		%bold-math for vectors

\begin{document}

\markboth{Yevgeny.~V.~Stadnik \and Victor.~V.~Flambaum}
{New Atomic probes for Dark Matter detection}

%%%%%%%%%%%%%%%%%%%%% Publisher's Area please ignore %%%%%%%%%%%%%%
%\catchline{}{}{}{}{}
%%%%%%%%%%%%%%%%%%%%%%%%%%%%%%%%%%%%%%%%%%%%%%%%%%%%%%%%%%%%%%%%%%%

\title{New Atomic probes for Dark Matter detection: 
\\Axions, Axion-like particles and Topological Defects  
}

\author{Yevgeny V.~Stadnik \and Victor V.~Flambaum}

\address{School of Physics, University of New South Wales, Sydney 2052, Australia
\\y.stadnik@unsw.edu.au}

\maketitle

%\pub{Received (Day Month Year)}{Revised (Day Month Year)}

\begin{abstract}
We present a brief overview of recently proposed detection schemes for axion, axion-like pseudoscalar particle and topological defect dark matter. We focus mainly on the possibility of using atomic and molecular systems for dark matter detection. For axions and axion-like particles, these methods are complementary probes to ongoing photon-axion interconversion experiments and astrophysical observations. For topological defects, these methods are complementary to conventional astrophysical search schemes based on gravitational signatures.

\end{abstract}

\ccode{PACS Nos.: 95.35.+d, 14.80.Va, 95.36.+x, 11.27.+d, 07.55.Ge, 06.30.Ft, 11.30.Er, 06.20.Jr}

%%%%%%%%%%%%%%%%%%%%%%%
\section{Introduction}	
Cold dark matter remains one of the most important unsolved problems in contemporary physics. The matter-energy content of the universe is overwhelming dominated by cold dark matter ($\sim 23 \%$) and dark energy ($\sim 73 \%$), with only a few percent attributable to baryonic matter (see e.g.~Ref.~\cite{Spergel07}). However, the composition and properties of both cold dark matter and dark energy still remain unclear. Some of the most popular candidates for dark matter include weakly interacting massive particles (WIMPs), super-weakly interacting massive particles (super-WIMPs), massive astrophysical compact halo objects (MACHOs), such as primordial black holes, and axions (see e.g.~Ref.~\cite{PDG12} and the plethora of references therein for further details of the properties of and search efforts for these particles). 

The strong $\mathcal{CP}$ problem, which seeks to explain why quantum chromodynamics (QCD) does not appear to violate the combined charge-parity ($\mathcal{CP}$) symmetry, is also an outstanding problem of great importance. The QCD Lagrangian contains the $\mathcal{P}$,$\mathcal{CP}$-violating term:
\begin{equation}
\label{QCD-theta}
\mathcal{L}^{\theta}_{\textrm{QCD}} = \theta \frac{g^2}{32 \pi^2} G\tilde{G} ,
\end{equation}
where $\theta$ is the angle quantifying the amount of $\mathcal{CP}$ violation within the QCD sector, $g^2/4\pi$ is the color coupling constant, and $G$ and $\tilde{G}$ are the gluonic field tensor and its dual, respectively. The observed value of $\theta$ from measurements of the permanent neutron electric dipole moment (EDM) is presently $|\theta| < 10^{-10}$ \cite{Baker06-NEDM}. The smallness of this measured angle, which may in principle assume any value in the range $-\pi \le \theta \le \pi$, constitutes the strong $\mathcal{CP}$ problem.

%%One possible resolution of the strong $\mathcal{CP}$ problem is that the QCD $\mathcal{CP}$-symmetry breaking parameter $\theta$ becomes unobservable if at least one of the quarks is massless (see e.g.~Ref.~\cite{Kim_RMP_10}). However, there does not appear to be empirical evidence to date that any of the quarks in the Standard Model (SM) are massless and so this resolution mechanism seems unlikely. 

An elegant explanation of the strong $\mathcal{CP}$ problem is offered by the Peccei-Quinn (PQ) theory, in which an additional global U($1$) symmetry, known as the PQ symmetry, is introduced into the SM QCD Lagrangian and is subsequently broken both spontaneously and explicitly \cite{PQ77a,PQ77b}. See also Refs.~\cite{Weinberg78x,Wilczek78x,Kim79x,Zakharov80x,Zhitnitsky80x,Srednicki81x}. The breaking of the PQ symmetry gives rise to a pseudoscalar pseudo-Nambu-Goldstone boson, known as the axion, being born from the QCD vacuum and causes the $\theta$ parameter to become effectively zero. It has been noted that the axion may also be a promising CDM candidate. Thus axions, if detected, could resolve both the dark matter and strong $\mathcal{CP}$ problems \cite{Kim_RMP_10,Kawasaki2013,Brambilla2014,Baer2014}. The decay of supersymmetric axions to produce axions has also been suggested as a possible explanation for dark radiation \cite{Jeong2012,Graf2013a,Graf2013b,Queiroz2014}.

%%ALPs--- not related to QCD axion

Most of the main Earth-based detection schemes for axions are based on the axion-photon coupling:
\begin{equation}
\label{FF-tilde-axion}
\mathcal{L}_{a \gamma \gamma} = - \frac{g_{a \gamma \gamma}}{4} a F\tilde{F} ,
\end{equation}
where $g_{a \gamma \gamma}$ is the coupling constant for the given interaction, $a$ is the axion field, and $F$ and $\tilde{F}$ are the photon field tensor and its dual, respectively. Haloscope searches aim to detect the electromagnetic power generated in the conversion of axions to photons inside microwave cavities \cite{Sikivie1983,Wuensch1989,Hagmann1990,Yamamoto2001,Asztalos2002,ADMX2014}. Helioscope searches aim to detect solar axions through their conversion into photons inside a strong magnet, which is directed toward the Sun \cite{Sikivie1983,Tokyo1998,Tokyo2002,CAST2007,Tokyo2008,Tokyo2013,CAST2014,IAXO2014}. A variety of photon regeneration experiments in the form of either light-shining-through-a-wall or vacuum birefringence effects have also been proposed and performed \cite{Anselm1985,Bibber1987,Hoogeveen1992,Rabadan2006,Chou2008,PVLAS2008,Caspers2009,Dias2009,Chou2009,Ehret2010,Battesti2010,Redondo2011,Baehre2013,Betz2013,OSQAR2013}. Recently, axion-like-particles have been suggested as a possible explanation for the anomalous 3.5 keV X-ray line in observations of the Andromeda galaxy \cite{Bulbul2014,Boyarsky2014,Higaki2014,Park2014,Lee2014,Takahashi2014}. For a recent overview of ongoing and near-future experiments, we refer the reader to Ref.~\cite{Ringwald2014}.

Axions and axion-like particles can also interact with the axial-vector currents of nucleons and the electron via the derivative-type interaction:
\begin{equation}
\label{DT-coupling}
\mathcal{L}_{a XX} = - \frac{C_X}{f_a} \left(\partial_\mu a\right)  \left(\bar{\psi}_X \gamma^\mu \gamma^5 \psi_X \right) ,
\end{equation}
where $X=e,p,n$ denotes the fermion species, $f_a$ is the axion decay constant and $\bar{\psi}_X \gamma^\mu \gamma^5 \psi_X$ is the axial-vector current associated with the fermion. $C_X$ is a model-dependent constant, which is typically taken to be $\sim 1$ for all $X=e,p,n$. A common axion detection scheme based on the interaction (\ref{DT-coupling}) involves the axio-electric effect, which is the ionisation of (usually atomic) matter by axions (as opposed to by photons in the photo-electric effect)~\cite{Avignone1987,Pospelov2008,Avignone2009a,Avignone2009b,Derevianko2010,Dzuba2010,Derbin2012,Derbin2013}. Bounds obtained from astrophysical data assist in axion searches by ruling out large regions of the allowed values of axion parameters \cite{Dicus78,Khlopov1978,Raffelt1986,Ellis1987,Mayle1988,Turner1988,Kolb1989,Engel1990,Isern1992,Raffelt1995,Janka1996,Raffelt08,Isern2008,Isern2012}.

Apart from the more usual variants of dark matter, non-trivial, stable configurations of scalar, vector or axion-like pseudoscalar fields, which may have formed at early cosmological times, are also possible. Such dark matter configurations are generally termed as topological defect dark matter (or more simply as `defects') and can have various dimensionalities: 0D (corresponding to monopoles), 1D (strings) and 2D (domain walls) \cite{Kibble1976,Zeldovich1980,Vilenkin1981,Lazarides1982,Sikivie1982defects,Vilenkin1982,Rubakov1983,Vilenkin1983,Vilenkin1985PR}. The transverse size $d$ of a defect cannot be predicted from existing theory in an \emph{ab initio} manner, but typically scales as $d \propto 1/m_\phi$, where $m_\phi$ is the mass of the particles making up the defect. Defects may contribute to the total dark matter content of the universe. These objects have primarily been sought for via their gravitational effects, including gravitational lensing and gravitational radiation \cite{Hogan1984,Vilenkin1984,Vachaspati1985,Blandford1992,Caldwell1992,TDDM_Book1,Brunstein1995,Caldwell1996,deLaix1997,TDDM_Book2,Maggiore2000,Damour2000,TDDM_Book3,Sazhin2003,Sazhin2006,Hubble2006,Siemens2006,Mack2007,Siemens2007,LIGO2009}.  Constraints on the contribution of cosmic defects to observed temperature fluctuations in the cosmic microwave background radiation spectrum have been placed by recent results from the Planck \cite{Planck2013-Defects} and BICEP2 collaborations \cite{BICEP2014,Moss2014-BICEP2}, but the existence of cosmic defects is neither confirmed nor ruled out by these results, leaving the question of whether or not defects exist largely unanswered.

Atomic systems have been employed with great success as high precision frequency standards \cite{DereviankoKatori2011Rev}, while atomic and molecular systems have provided sensitive tests of the SM and have served as sensitive probes of searches for physics beyond the SM \cite{Khriplovich1991_PNC-Book,Kostelecky1999,Ginges_PhysRep2004,PospelovRitz2005,Kostelecky2011data,DZUBA2012-Review(PNC),Engel2013}. Atomic clocks are arguably the most precise instruments ever built by mankind, with the best current fractional inaccuracies of the order $\sim 10^{-18}$ \cite{Hinkley2013,Bloom2014}. At present, atomic Hg provides the most precise limits on the EDM of the proton, quark chromo-EDM and $\mathcal{P}$,$\mathcal{T}$-odd nuclear forces, as well as the most precise limits on the neutron EDM and QCD $\theta$ term from atomic or molecular experiments \cite{Griffith2009improved,Swallows2013}, while molecular ThO provides the most precise limit on the electron EDM \cite{Baron2014}. Measurements and calculations of the $6s$-$7s$ parity nonconserving (PNC) amplitude in atomic Cs stand as the most precise atomic test of the SM electroweak theory to date, see e.g.~Refs.~\cite{Bouchiat1982Cs,Wood1997_Cs-PNC,Dzuba1989Cs,Blundell92Cs,Kozlov01Cs,Safronova02Cs,Ginges02Cs,Roberts2012_Cs-pnc}, and are competitive with direct searches performed at hadron colliders \cite{PDG12,Abulencia2006}. Atomic co-magnetometers provide some of the most stringent limits on $\mathcal{CPT}$- and Lorentz-invariance-violating physics \cite{Berglund1995,Kostelecky1999,Cane2004,Gemmel2010,Brown2010,Kostelecky2011RMP,Peck2012,Allmendinger2014,Stadnik-NMBE2014}. 

The possibility of using atomic systems as highly sensitive probes for axions, axion-like particles and topological defects has only recently been explored. In the following sections, we present a brief overview of recent proposals for: 

(I) The detection of axions and axion-like pseudoscalar particles using atomic, molecular and solid-state systems.

(II) The detection of topological defects using atomic, molecular and astrophysical systems via non-gravitational signatures.

%%%%%%%%%%%%%%%%%%%%%%%
\section{Axions and axion-like pseudoscalar particles}	
It was pointed out in Refs.~\cite{Graham2011,Graham2013} that due to the coherently oscillating nature of the axion field, axions induce oscillating nucleon EDMs via the interaction (\ref{QCD-theta}). In the case of a free neutron, the axion-induced EDM is given by \cite{Pospelov-Ritz1999,Graham2011,Graham2013}: 
\begin{equation}
\label{NEDM}
d_n = 2.4 \times 10^{-16} \theta ~ e \cdot \textrm{cm} = 2.4 \times 10^{-16} \frac{a}{f_a} e \cdot \textrm{cm} .
\end{equation}
These oscillating nucleon EDMs give rise to oscillating $\mathcal{P}$,$\mathcal{T}$-odd electromagnetic moments in nuclei \cite{Graham2011,CASPEr2014,Stadnik2014Axion,Roberts2014in-prep}. For most nuclei of experimental interest, the dominant mechanism for generating $\mathcal{P}$,$\mathcal{T}$-odd nuclear electromagnetic moments is via the $\mathcal{P}$,$\mathcal{T}$-violating nucleon-nucleon interaction mediated by pion exchange (the source of $\mathcal{P}$,$\mathcal{T}$-violation is provided by the axion field) \cite{Stadnik2014Axion}. In the quasi-static approximation, in which the axion mass energy is much smaller than the energy separation between the atomic state of interest and all opposite-parity states, the lowest-order $\mathcal{P}$,$\mathcal{T}$-odd nuclear electromagnetic moment -- the nuclear EDM -- is fully screened as a direct consequence of Schiff's theorem, \cite{Schiff196x} and the dominant contribution to the $\mathcal{P}$,$\mathcal{T}$-odd nuclear electromagnetic moment arises from the nuclear Schiff moment \cite{Flambaum1984PT,Flambaum1986,Stadnik2014Axion}. 

One method of detecting axion-induced oscillating nuclear Schiff moments is via precision magnetometry performed on a sample of pre-polarised nuclear spins in a solid-state medium in the presence of static electric and magnetic fields \cite{CASPEr2014}. In this type of experiment, matching of the Larmor frequency to the axion mass energy results in the precession of the nuclear spins, in a manner similar to nuclear magnetic resonance techniques. The induced transverse magnetisation can then be detected with a SQUID pickup loop. Static EDM experiments in ferroelectrics have been proposed and performed \cite{Lamoreaux2006,Lamoreaux2012}. Axion dark matter detection schemes employing Josephson junctions \cite{Beck2014a,Beck2014b} and LC circuits \cite{Sikivie2013LC} have also been proposed recently. 

Another method of detecting oscillating nuclear Schiff moments is via the oscillating EDMs they induce in atoms and molecules \cite{Stadnik2014Axion}. Axions also induce oscillating $\mathcal{P}$,$\mathcal{T}$-odd effects in molecules via interaction (\ref{QCD-theta}) through the generation of oscillating nuclear magnetic quadrupole moments (MQMs), which arise from $\mathcal{P}$, $\mathcal{T}$-odd intranuclear forces and from the EDMs of constituent nucleons \cite{Roberts2014in-prep}. Unlike nuclear EDMs, nuclear MQMs are not screened by the atomic electrons \cite{Flambaum1984PT,Flambaum2014MQM}. The temporal component in the derivative-type interaction of the form (\ref{DT-coupling}) between axions and electrons induces oscillating EDMs and PNC effects in atoms and molecules, while the temporal component in the derivative-type interaction (\ref{DT-coupling}) between axions and nucleons induces oscillating nuclear anapole moments \cite{Stadnik2014Axion,Roberts2014Cosmic-pnc(prl),Roberts2014in-prep}. Nuclear anapole moments manifest themselves as a nuclear-spin-dependent contribution to a PNC transition amplitude in atoms and molecules \cite{Flambaum1984NAM}.

The spatial components in the derivative-type interaction (\ref{DT-coupling}) give rise to the coupling $\v{s} \cdot \v{p}_a \cos(m_a t)$ (also known as the axion-wind effect) between the spin angular momentum of a fermion, $\v{s}$, and the axion linear momentum, $\v{p}_a$ \cite{AW1995a,AW1995b,Graham2013,Stadnik2014Axion}. This coupling is of the same form as that for the interaction of a spin with an oscillating magnetic field, $\v{s} \cdot \v{B} \cos(\omega t)$, and can be sought for through spin precession experiments. Since the effective magnetic field strength due to this interaction is inversely proportional to the magnetic dipole moment of the fermion species considered, nuclear spin precession experiments are preferable to electron spin precession experiments. The axion field is modified in the presence of Earth's gravitational field. The interaction of the spin angular momentum of a fermion with this modified axion field leads to an oscillating axion-induced spin-gravity coupling of the form $\v{s} \cdot \v{g} \cos(m_a t)$, where $\v{g}$ is the gravitational field direction \cite{Stadnik2014Axion}. Static spin-gravity couplings have been considered and sought for previously (see e.g.~Refs.~\cite{Venema1992,FLP2009,Kimball2013}). A variety of co-magnetometers may be used to search for such spin-dependent couplings \cite{Romalis2011Ne,Peck2012,Kimball2013,Romalis2014Rb-Xe-He-new}. Consideration of nuclear many-body effects, such as polarisation of the nuclear core by the valence nucleon(s), is important for interpretations of axion searches through nuclear-spin-dependent effects, including $\mathcal{P}$,$\mathcal{T}$-odd effects generated through hadronic mechanisms, spin-axion momentum and spin-gravity couplings \cite{Stadnik-NMBE2014}.

%Axion-induced effects, which arise from the derivative-type coupling (\ref{DT-coupling}), share some similarities with couplings of the form
%\begin{equation}
%\label{DT-coupling}
%\mathcal{L}_{int} = - b_\mu^X  \left(\bar{\psi}_X \gamma^\mu \gamma^5 \psi_X \right) ,
%\end{equation}
%between a time-dependent pseudovector field, $b_\mu$, and the axial-vector current associated with a fermion. Static pseudovector fields with the coupling (\ref{DT-coupling}) have been considered in the context of $\mathcal{CPT}$- and Lorentz-invariance-violating theories beyond the Standard Model \cite{Kostelecky1997,Kostelecky1998}. Constraints on the spatial components of these fields have been obtained through experiments searching for sidereal modulations of Zeeman transitions frequencies in co-magnetometers \cite{Berglund1995,Kostelecky1999,Cane2004,Gemmel2010,Brown2010,Peck2012,Allmendinger2014,Stadnik-NMBE2014}, while direct constraints on the temporal components of such fields have been obtained from considerations of PNC transition amplitudes \cite{Roberts2014Cosmic-pnc(prl),Roberts2014in-prep,Stadnik-NMBE2014}. 

%%Add Kam/Prad ref. ~ searches for low-mass axions (oscillating nuclear Schiff moments)... ???

%%%%%%%%%%%%%%%%%%%%%%%
\section{Topological defects}	
Stable domain walls consisting of QCD-type axions could lead to disastrous cosmological consequences by storing too much energy \cite{Sikivie1982defects}. However, topological defects, which consist of axion-like pseudoscalar particles, are possible and may interact with fermions via the linear-in-$a$ derivative-type interaction (\ref{DT-coupling}) and a corresponding quadratic-in-$a$ interaction [with $a/f_a \to (a/f'_a)^{2}$ in Eq.~(\ref{DT-coupling})]. The spatial components in these interactions give rise to an interaction of the form $\v{s} \cdot (\v{\nabla}a) \equiv \v{\mu} \cdot \v{B}_{\textrm{eff}}$, where $\v{\mu}$ is the fermion magnetic dipole moment, \cite{Budker2012Wall} while the temporal components in these interactions give rise to transient EDMs \cite{Stadnik2014defects}. Both of these effects can be sought for with a synchronised global network of magnetometers (GNOME) \cite{GNOME2013}. Account of nuclear many-body effects is important in such searches \cite{Stadnik-NMBE2014,Kimball2014NSD}.

Topological defects, which consist of scalar fields, may alter the fermion masses via the coupling:
\begin{equation}
\label{L_Varn_FCs}
\mathcal{L}_{\textrm{int}} = - \sum_{X=e,p,n} m_X \left(\frac{\phi}{\Lambda_X}\right)^2 \bar{\psi}_X \psi_X ,
\end{equation}
where $\phi$ is the scalar field, $\bar{\psi}_X \psi_X$ is the scalar density associated with the fermion field, $m_X$ is the standard mass of the fermion and $\Lambda_X$ is the reciprocal of the coupling constant for the given interaction with a particular fermion $X$. The passage of a defect through Earth can be detected with a synchronised global network of atomic clocks, which become temporarily desynchronised (atomic hyperfine, as well as molecular hyperfine, rotational and vibrational transition frequencies are sensitive to alterations in the fermion masses) upon the passage of a defect through Earth \cite{Derevianko2013Defect-clock}. A large variety of hyperfine transitions in atomic species \cite{Flambaum2006A,Dinh2009,Berengut2011a,Guena2012}, as well as hyperfine, rotational and vibrational transitions in molecular species \cite{Flambaum2006C,Flambaum2007D,DeMille2008A,Zelevinsky2008A,Kozlov2009A,Kozlov2013B,Kozlov2013C,Flambaum2013B} can be used to search for variations of neutron, proton and electron masses, induced by interaction (\ref{L_Varn_FCs}). Account of nuclear many-body effects is important in such searches \cite{Flambaum2006A,Berengut2011}. Temporary alterations in fermion masses would also alter Earth's period of rotation upon defect passage through Earth and these alterations could be measured by monitoring Earth's angle of rotation over time using an atomic clock \cite{Stadnik2014defects}. A large variety of atomic systems can be used for this purpose \cite{Allan1997,Capitaine2001,Peik2004,Takamoto2005opticalSrLAttice,Oskay2006,Tobar2006Sapphire,Ludlow2008,Chou2010,Jiang2011Yb_opt_lattice,DereviankoKatori2011Rev,Hinkley2013,Bloom2014}.

Defects may also induce alterations in pulsar rotational frequencies by temporarily altering the mass, radius and possibly the internal structure of a pulsar upon their passage through a pulsar via interaction (\ref{L_Varn_FCs}) \cite{Stadnik2014defects}. The photon mass might be non-zero inside a defect object and so defect objects might also induce time-delay of signals from pulsars by passing between the line-of-sight connecting a pulsar and Earth, as well as lens background radiation in a manner distinct to a gravitational lens (in the presence of a non-zero photon mass, the index of refraction inside a defect would scale as $n(\omega) \approx 1 + m_\gamma^2 / (2\omega^2)$, where $m_\gamma$ is the photon mass inside a defect) \cite{Stadnik2014defects}.

\section*{Acknowledgments}

This work was supported by the Australian Research Council. V.~V.~F.~would also like to acknowledge the Humboldt foundation for support and the MBN Research Center for hospitality.

\end{document}